\documentstyle[preprint,aps,pra,epsfig,cite]{revtex}
\begin{document}

\draft
\title{Entangled Nonorthogonal States and
Their Decoherence Properties
}
\author{Osamu Hirota$^{1,2}$, Steven J. van Enk$^+$,
Kazuo Nakamura$^{\ast1,2}$,\\ Masaki Sohma$^\dagger$,
and Kentaro Kato$^{1,2}$\\
1.Tamagawa University, Machida, Tokyo, Japan.\\
2. CREST, JST$:$Japan Science and Technology.\\
$^+$Bell Labs, Lucent Technologies, Murray Hill, NJ, USA\\
$^\ast 1$ Fundamental Research Laboratory, NEC, Tsukuba, Japan.\\
$^\ast 2$ CREST, JST$:$Japan Science and Technology.\\
$^\dagger$ Matsushita Research Institute Inc. Tokyo, Japan.
}

\maketitle

\begin{abstract}
This paper presents properties of the so-called quasi-Bell states: 
entangled states written as superpositions of nonorthogonal states.
It is shown that a special class of those states, namely entangled 
coherent states, are more robust against decoherence due to 
photon absorption than the standard bi-photon Bell states.
\end{abstract}

\pacs{PACS numbers: 03.65.Bz}

\section{Introduction}
Entanglement and its information-theoretic aspects have been
studied by many authors
\cite{Bennett,Bennett2,vedral,Wootters,Wootters2}.
Here we give a short survey of the theory of entanglement that 
we will later apply to quasi-Bell states.
For a pure entangled state of a bipartite system
$|\Psi\rangle_{AB}$, the measure of entanglement defined as
\cite{Bennett,Barnett}
\begin{equation}
E(|\Psi\rangle_{AB})=-{\rm Tr}_{A}{\rho}_{A}\log {\rho}_{A},
\quad
{\rho}_{A}={\rm Tr}_{B}|\Psi\rangle_{AB}\langle\Psi|,
\end{equation}
is called the ``entropy of entanglement''.
This quantity enjoys two kinds of information-theoretic
interpretations.
One is that $E$ gives the entanglement of formation, which is defined as
the asymptotic number $k$ of standard singlet states required to
faithfully locally prepare  $n$ identical copies of
a system in the bipartite state $|\Psi\rangle_{AB}$ for
very large $k$ and $n$.
The other is that $E$ gives the amount of distillable entanglement, 
which is the asymptotic number of singlets $k$
that can be distilled from $n$ identical copies of
$|\Psi\rangle_{AB}$.
With either of these definitions of $k$ and $n$, $E$ satisfies
\begin{equation}
\lim _{n, k \rightarrow \infty} \frac{k}{n} = E(|\Psi\rangle_{AB}).
\end{equation}
For pure states we can rewrite
\begin{equation}
E(|\Psi\rangle_{AB})=
H\left(\frac{1}{2}(1+\sqrt {1 - C(|\Psi\rangle_{AB})^2})\right)
\end{equation}
where $H(x) = -x \log x -(1-x)\log (1-x)$ is the binary entropy function
and
$C(|\Psi\rangle_{AB})$ is the ``concurrence'' defined by
$C(|\Psi\rangle_{AB})=
|_{AB}\langle\Psi|\tilde\Psi\rangle_{AB}|$ with
$|\tilde\Psi\rangle_{AB}=\sigma|\Psi\rangle_{AB}^*$. The above expression 
is valid for mixed states of two qubit systems as well \cite{Wootters}.

For mixed states of qubits one may also define
an expression for the entanglement of formation
\cite{Bennett,Wootters,Wootters2}.
It is defined as the average entanglement of
the pure states of a decomposition 
$\rho=\sum_ip_i|\psi\rangle_i\langle\psi_i|$
of the density operator $\rho$, minimized over
all decompositions \cite{Bennett}
\begin{equation}
E(\rho)= \min \sum p_i E(|\psi_i \rangle).
\end{equation}
In general, it is difficult to find the exact amount 
of entanglement of formation except for special cases.
However, there is a lower bound which is expressed
in terms of a quantity called the ``fully entangled fraction'', 
which we denote by $f(\rho)$ and is defined as
\begin{equation}
f(\rho)=\max \langle e |\rho | e \rangle,
\end{equation}
where the maximum is over all completely entangled states
$ | e \rangle $.
A lower bound on the entanglement of formation is \cite{Bennett}
\begin{equation}
E(\rho) \ge h[f(\rho)],
\end{equation}
where
\begin{equation}
h[f(\rho)] = \left\{
\begin{array}{ll}
H[ \frac{1}{2} + \sqrt{f(1-f)}] & (f \ge \frac{1}{2})\\
0  & (f < \frac{1}{2})\\
\end{array}
\right.
\end{equation}
Usually in order to construct entangled states one writes 
superpositions of orthogonal states. 
For instance the standard Bell basis uses states like 
$|\updownarrow\rangle$ and $|\leftrightarrow\rangle$, 
and of course its properties are well known.
Our concern here is what kind of properties appear if we 
have superpositions of nonorthogonal states.
In this paper, we will clarify properties of entangled states
of nonorthogonal states such as coherent states based on the above
basic theory.

\section{Quasi-Bell states}
\subsection{General definition}
Let us consider entangled states based on two nonorthogonal states
$|\psi_1 \rangle$ and $|\psi_2 \rangle$ such that
$ \langle \psi_1 | \psi_2 \rangle = \kappa$ where $\kappa$ is real. 
We can define a set of 4 entangled states as follows:
\begin{eqnarray}
\left\{
\begin{array}{lcl}
|\Psi_1 \rangle_{AB} &=& h_{1} (|\psi_1 \rangle_A|\psi_2 \rangle_B
+|\psi_2 \rangle_A|\psi_1 \rangle_B ) \\
|\Psi_2 \rangle_{AB} &=& h_{2} (|\psi_1 \rangle_A|\psi_2 \rangle_B
-|\psi_2 \rangle_A|\psi_1 \rangle_B )\\
|\Psi_3 \rangle_{AB} &=& h_{3} (|\psi_1 \rangle_A|\psi_1 \rangle_B
+ |\psi_2 \rangle_A|\psi_2 \rangle_B )\\
|\Psi_4 \rangle_{AB} &=& h_{4} (|\psi_1 \rangle_A|\psi_1 \rangle_B
-|\psi_2 \rangle_A|\psi_2 \rangle_B)
\end{array}
\right.
\end{eqnarray}
where
$\{h_{i}\}$ are normalization constants:
$h_{1}=h_{3}=1/\sqrt{2(1+\kappa^{2})}$,
$h_{2}=h_{4}=1/\sqrt{2(1-\kappa^{2})}$.
We call these states  ``quasi-Bell states''.
They are not orthogonal to each other.
In fact for $\kappa$ real their Gram matrix
$G_{ij}=|_{AB}\!\langle\Psi_i|\Psi_j\rangle_{AB}|$ becomes 
\begin{equation}
G=
\left(
\begin{array}{cccc}
1& 0& D& 0\\
0& 1& 0& 0\\
D& 0& 1& 0\\
0& 0& 0& 1\\
\end{array}
\right)
\label{ohmsgrammat}
\end{equation}
where $D=2 \kappa/(1 + {\kappa}^2)$.
If the basic states are orthogonal ($\kappa=0$), then these states 
reduce to standard Bell states.
Let us discuss the entropy of entanglement for the above states.
We first calculate the reduced density operators of
the quasi-Bell states.
They are ${{\rho}_{A}}^{(1)}={{\rho}_{A}}^{(3)}$ and
${{\rho}_{A}}^{(2)} = {{\rho}_{A}}^{(4)} $ with
\begin{equation}
{{\rho}_{A}}^{(1)}
 = \frac{1}{2(1+\kappa^2)}
\{|\psi_1 \rangle\langle\psi_1| +
\kappa |\psi_1 \rangle\langle \psi_2|
+\kappa |\psi_2 \rangle\langle \psi_1|
+ |\psi_2 \rangle\langle \psi_2\},
\end{equation}
\begin{equation}
{{\rho}_{A}}^{(2)}=\frac{1}{2(1-\kappa^2)}
\{|\psi_1 \rangle_A\langle \psi_1|
-\kappa |\psi_1 \rangle_A\langle \psi_2|
-\kappa |\psi_2 \rangle_A\langle \psi_1|
+ |\psi_2 \rangle_A\langle \psi_2|\}.
\end{equation}
The eigenvalues of the above density operators
${{\rho}_{A}}^{(1)}$(or ${{\rho}_{A}}^{(3)}$) are given
in terms of the Gram matrix elements $G_{ij}$ as follows,
\begin{equation}
\lambda_{1/1} = \frac{1+G_{13}}{2}=
\frac{(1+\kappa)^2}{2(1+\kappa^2)},
\quad
\lambda_{2/1} = \frac{1-G_{13}}{2}
=\frac{(1-\kappa)^2}{2(1+\kappa^2)},
\end{equation}
and for ${{\rho}_{A}}^{(2)}$(or ${{\rho}_{A}}^{(4)}$)
we have
\begin{equation}
\lambda_{1/2}= \frac{1+G_{24}}{2}=\frac{1}{2},
\quad
\lambda_{2/2}= \frac{1-G_{24}}{2}=\frac{1}{2}.
\end{equation}
Hence, the entropy of entanglement is
\begin{equation}
E(|\Psi_1\rangle_{AB})=E(|\Psi_3\rangle_{AB})
                 = - \frac{1+D}{2} \log \frac{1+D}{2}
                   - \frac{1-D}{2} \log \frac{1-D}{2},
\end{equation}
and
\begin{equation}
E(|\Psi_2\rangle_{AB})=E(|\Psi_4\rangle_{AB})=1,
\end{equation}
because $G_{13}=D$, and $G_{24}=0$.
Thus $|\Psi_2\rangle_{AB}$ and $|\Psi_4\rangle_{AB}$
are maximally entangled, even though the entangled states
consist of nonorthogonal states in each subsystem.
These results are true for arbitrary nonorthogonal states
with $\langle \psi_1 | \psi_2 \rangle$ $=$
$\langle \psi_2 | \psi_1 \rangle$ $=$ $\kappa$ and
do not depend on the physical dimension of
the systems. This property may be unexpected but can be 
understood easily by noting that the states $|\Psi_2\rangle$
and $|\Psi_4\rangle$ are equivalent to 
$(|+\rangle|-\rangle \pm |-\rangle|+\rangle)/\sqrt{2}$ 
in terms of the orthogonal basis
\begin{equation}
|\pm\rangle=(|\psi_1\rangle \pm |\psi_2\rangle)/\sqrt{N_{\pm}},
\end{equation}
with $N_{\pm}=2\pm 2\kappa$.

\subsection{Mixtures of quasi Bell states}
We can construct a quasi-Werner mixed state based on 
quasi-Bell states by
\begin{equation}
W = F|\Psi_2 \rangle_{AB} \langle \Psi_2 | +
\frac{1-F}{3}\{|\Psi_1 \rangle_{AB}  \langle \Psi_1|
+|\Psi_3 \rangle_{AB} \langle \Psi_3 | +
|\Psi_4 \rangle_{AB} \langle \Psi_4 |\},
\end{equation}
where $0 \le F \le 1$.
If $|\Psi_1 \rangle_{AB}$,$|\Psi_2 \rangle_{AB}$,
$|\Psi_3 \rangle_{AB} $,
$|\Psi_4 \rangle_{AB}$ are Bell states, then
the above equation gives a standard Werner state \cite{werner}.
It is known that the fully entangled fraction of
the Werner state is $F$, and the entanglement of
formation of the Werner state is given by
\begin{equation}
E(W) = H\left(\frac{1}{2} + \sqrt{F[1-F]}\right).
\end{equation}
The fully entangled fraction of
the quasi Werner state is analogously given by
\begin{equation}
f(W) = _{AB}\!\langle \Psi_2 |W |\Psi_2 \rangle_{AB} = F,
\end{equation}
because the quasi Bell states are orthogonal to each other,
except for the pair of states
$|\Psi_1 \rangle_{AB}$ and $|\Psi_3 \rangle_{AB} $,
as one can see from the Gram matrix $G$.
However, the quasi Werner state and Werner state are
completely different states.
In particular, the eigenvalues of quasi Werner states are
different from those of Werner states.
The eigenvalues of the density operator are given by those of
the modified Gram matrix
\cite{hirota0}. For the quasi Werner state,
the Gram matrix is
\begin{equation}
G_W=
\left(
\begin{array}{cccc}
\frac{1}{3}(1-F)& 0& \frac{1}{3}(1-F)D& 0\\
0& F& 0& 0\\
\frac{1}{3}(1-F)D& 0& \frac{1}{3}(1-F)& 0\\
0& 0& 0& \frac{1}{3}(1-F)\\
\end{array}
\right)
\end{equation}
As a result, we have
$F$, $\frac{1}{3}(1-F)$, $\frac{1}{3}(1+D)(1-F)$,
$\frac{1}{3}(1-D)(1-F)$ as the eigenvalues of
the quasi Werner state.

Thus the lower bound of the entropy of formation of
quasi Werner state is the same as that of Werner state.
For more general mixtures of quasi Bell states
we will need a more advanced analysis which is
reported on in a subsequent paper.

\section{Quasi-Bell states based on bosonic coherent states}
Let us consider two coherent states of a bosonic mode
$\{|\alpha\rangle, |-\alpha\rangle\}$, e.g., let
$\pm \alpha$ be the coherent amplitude of a 
light field. Using previous notation, we have
$\kappa=\langle\alpha|-\alpha\rangle=\exp\{-2|\alpha|^2\}$.
Then one can construct the quasi Bell states as follows:
\begin{eqnarray}
\left\{
\begin{array}{lcl}
|\Psi_1 \rangle_{AB}  &=&
h_{1} (|\alpha \rangle_A|-\alpha \rangle_B
+|-\alpha \rangle_A|\alpha \rangle_B ) \\
|\Psi_2 \rangle_{AB}  &=&
h_{2} (|\alpha \rangle_A|-\alpha \rangle_B
-|-\alpha \rangle_A|\alpha \rangle_B )\\
|\Psi_3 \rangle_{AB}  &=&
h_{3} (|\alpha \rangle_A|\alpha \rangle_B
+ |-\alpha \rangle_A|-\alpha \rangle_B )\\
|\Psi_4 \rangle_{AB}  &=&
h_{4} (|\alpha \rangle_A|\alpha \rangle_B
-|-\alpha \rangle_A|-\alpha \rangle_B)
\end{array}
\right.
\end{eqnarray}
Since the coherent states are nonorthogonal,
we can apply the results of Section 2 to these states.
States of similar form were discussed by Sanders \cite{sanders},
and Wielinga \cite{wi}, who called these states entangled
coherent states. 
From the results in the Section 2, we know that 
$|\Psi_2 \rangle_{AB}$
and $|\Psi_4 \rangle_{AB}$ have
one ebit of entanglement independent of $\alpha$.
This is an interesting and potentially useful property 
(see next Section).
On the other hand, $|\Psi_1 \rangle_{AB} $ and
$|\Psi_3 \rangle_{AB}$ are
maximally entangled states only in the limit 
$\alpha\rightarrow\infty$. 
In order to avoid confusion with continuous
variable states \cite{kimble},
we should mention here that, of course, the dimension of
the space spanned by the quasi Bell states is 4 even though 
they are embedded in a vector space of infinite dimension.
This implies that the maximum value of
the von Neumann entropy for the quasi Bell states is unity.
Different amplitudes $\alpha$ just give the degree of nonorthogonality.
It also means that one cannot use coherent entangled states 
for teleportation of continuous quantum variables \cite{kimble,akira}.
In a separate paper \cite{enk}, we will show how to use 
entangled coherent states
for teleportation of Schr\"odinger cat states.
The average photon numbers of the reduced states of the
quasi Bell states read
\begin{equation}
\langle n_{A}^{(1)} \rangle =
\frac{(1-\kappa^2)}{(1+\kappa^2)} |\alpha |^2,
\quad
\langle n_{A}^{(2)} \rangle=
\frac{(1+\kappa^2)}{(1-\kappa^2)} |\alpha |^2.
\end{equation}
Thus the quasi Bell states can have arbitrary photon numbers.
As said above, however, the quasi Bell states are not truly 
continuous variable states and in particular
do not belong to the class of Gaussian states 
\cite{holevo},\cite{holevo2} in contrast to, e.g., 
the two mode squeezed state \cite{yuen}.
This is shown in the following way.
The characteristic functions of the quasi Bell states
are given by 
\begin{eqnarray}
C(\xi, \eta) &=& {\rm Tr}\big[
   |\Psi\rangle_{AB} \langle\Psi|
       \exp({\xi} {\mit a}_A^\dagger)
       \exp({-\xi^*} {\mit a}_A)
       \exp({\eta} {\mit a}_B^\dagger)
       \exp({-\eta^*} {\mit a}_B)
                           \big] \nonumber \\
&\times& \exp\{-(|\xi |^2 + |\eta |^2)/2\}
\end{eqnarray}
where $a$ and $a^\dagger$ are the annihilation and
creation operators, respectively.  
They can be calculated, with the result
\begin{eqnarray}
C(\xi, \eta |i=1, 2)&=&{h_{i}}^2\exp\{-(|\xi |^2 +
|\eta |^2)/2\}\{\exp(A_1-B_1)\alpha\nonumber\\
&+&\exp(-A_1+B_1)\alpha \pm \exp(A_2 - B_2)\alpha \nonumber \\
&\pm& \exp(-A_2 + B_2)\alpha\}\\
C(\xi, \eta |i=3,4)&=&{h_{i}}^2\exp\{-(|\xi |^2 +
|\eta |^2)/2\}\{\exp(A_1+B_1)\alpha\nonumber \\
&+&\exp(-A_1-B_1)\alpha \pm \exp(A_2 +B_2)\alpha \nonumber \\
&\pm& \exp(-A_2 - B_2)\alpha\}
\end{eqnarray}
where $A_1 = (\xi - \xi^*), A_2 = (\xi + \xi^*),
B_1 = (\eta - \eta^*),
B_2 = (\eta + \eta^*)$.
The characteristic functions are indeed not Gaussian.
Finally let us explore one more property of a similar set 
of entangled states.
If the amplitudes of the modes A and mode B
in $|\Psi_2 \rangle_{AB}$ (or $|\Psi_4 \rangle_{AB}$)
are chosen to be different, say, $\alpha$ and $\beta$, 
respectively, the eigenvalues of the reduced density operator
are
\begin{equation}
\lambda_{1} =
\frac{(1+\kappa_A)(1-\kappa_B)}{2(1-\kappa_A \kappa_B)},
\quad
\lambda_{2} =
\frac{(1-\kappa_A)(1+\kappa_B)}{2(1-\kappa_A \kappa_B)},
\end{equation}
where $\kappa_A=\exp(-2|\alpha|^2)$ and 
$\kappa_B=\exp(-2|\beta|^2)$.
We can then easily see that the entropy of entanglement 
attains its maximum value of 1
only when the amplitudes of both modes
are the same.

\section{Decoherence properties}
In this section, we will discuss 
decoherence properties of the state $|\Psi_2\rangle_{AB}$.
We are concerned with the decoherence due to energy loss or 
photon absorption.
In particular, we would like to demonstrate that entangled coherent 
states possess a certain degree of robustness against decoherence 
when compared to a standard bi-photon Bell state. 
We assume that
Alice produces a coherent entangled state $|\Psi_2\rangle_{AB}$, 
keeps one part ($A$) and transmits
the other part $B$ to Bob through a lossy channel. Bob will receive
an attenuated optical state. 
Thus, Alice prepares 
\begin{equation}
|\Psi_2\rangle_{AB}
=h_2( |\alpha\rangle_A |-\alpha\rangle_B
-|-\alpha\rangle_A |\alpha \rangle_B ) \equiv|\Psi_2(\alpha)\rangle_{AB}
\end{equation}
where $h_2=1/\sqrt{2(1-{\kappa_A}^2)}$,
$\kappa_A = \langle \alpha | -\alpha \rangle$.
When we employ a half mirror model for the noisy channel,
the effect of energy losses is described by a linear coupling
with an external vacuum field as follows:
\begin{equation}
U_{BE}|\alpha\rangle_B|0\rangle_E =
    |\sqrt{\eta}\alpha\rangle_B|\sqrt{(1-\eta)}\alpha\rangle_E.
\end{equation}

where the mode $E$ is an external mode responsible for the energy loss, 
$\eta$ is the noise parameter, and $\alpha$ is taken as real.
If we use $|\Psi_2(\alpha)\rangle_{AB} $ as the initial state,
the final state entangled with the environment is
\begin{eqnarray}
&\hat I_A\otimes U_{BE}|\Psi_2(\alpha)\rangle_{AB}
\otimes |0\rangle_E & \nonumber\\
&=h_2(|\alpha\rangle_A
|-\sqrt{\eta}\alpha\rangle_B|-\sqrt{(1-\eta)}\alpha\rangle_E
 -|-\alpha\rangle_A|\sqrt{\eta}\alpha\rangle_B|\sqrt{(1-\eta)}\alpha\rangle_
E ).&
\end{eqnarray}
The normalized density operator shared by Alice and Bob is given by
a super operator calculation \cite{barnett2},
\begin{eqnarray}
\rho_{AB}&=& h_2^2\{|\alpha \rangle \langle
\alpha |
\otimes |-\sqrt{\eta}\alpha \rangle \langle -\sqrt{\eta}\alpha |
+ |-\alpha \rangle \langle -\alpha |
\otimes |\sqrt{\eta}\alpha \rangle \langle \sqrt{\eta}\alpha |\nonumber \\
&-& |-\alpha \rangle \langle \alpha |
\otimes L |\sqrt{\eta}\alpha \rangle \langle -\sqrt{\eta}\alpha |
-|\alpha \rangle \langle -\alpha |
\otimes L |-\sqrt{\eta}\alpha \rangle \langle \sqrt{\eta}\alpha |\}
\end{eqnarray}
where
$L = \exp {\{-2(1-\eta)|\alpha|^2\}}$.
Let us discuss the entanglement of the above density
operator. As discussed in Section 2, we can use the entangled
fraction to measure the entanglement of this mixed state.
The fully entangled fraction of $\rho_{AB}$ is given by
\begin{equation}
f(\rho_{AB}) =\max_{\beta} \mbox{}_{AB}\! \langle \Psi_2(\beta)
|\rho_{AB}|\Psi_2(\beta)\rangle_{AB}
\end{equation}
because $|\Psi_2(\beta) \rangle_{AB}$ is indeed maximally entangled.
The above is given by
\begin{equation}
f(\rho_{AB})=\max_{\beta}\frac{(1+L)({\kappa_1}^2(\beta){\kappa_2}^2(\beta) + 
{\kappa_3}^2(\beta) {\kappa_4}^2(\beta)
-2 \kappa_1(\beta) \kappa_2(\beta) \kappa_3(\beta)\kappa_4(\beta))}
{2(1-{\kappa_A}^2)(1-{\kappa_0}^2(\beta))}
\end{equation}
where
$\kappa_0(\beta) = \exp \{-2|\beta|^2\}$, 
$\kappa_1(\beta) = \exp \{-|\alpha - \beta|^2/2\}$,
$\kappa_2(\beta) = \exp \{-|\beta - \sqrt{\eta}\alpha|^2/2\}$,
$\kappa_3(\beta) = \exp \{-|\alpha + \beta|^2/2\}$, and 
$\kappa_4(\beta) = \exp \{-|\beta + \sqrt{\eta}\alpha|^2/2\}$.
The maximum is attained for the value 
\begin{equation}
\beta=\frac{\alpha+\sqrt{\eta}\alpha}{2}
\end{equation}
that is, exactly halfway between the original coherent amplitude 
and the attenuated amplitude.
In Figure 1 we plot $f(\rho_{AB})$ as a function of $|\alpha|$ 
for various values of the noise parameter $\eta$. 
%%%%%%%%%%%%%%%%%%%%%%%%%%%%%%%%%%%%%%%%%
\begin{figure}[h]\label{f1} \leavevmode
\centerline{
	\epsfig{file=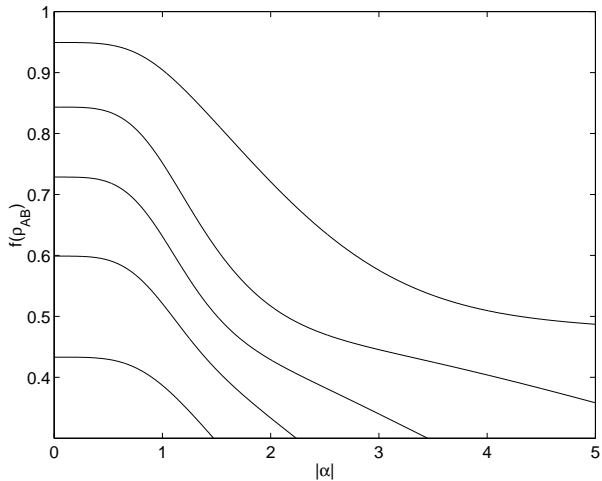,width=8cm}
	%\epsfile{file=ent.eps,width=8cm}
}
\caption{Overlap of $\rho_{AB}$ 
with appropriately chosen fully entangled state (see text) as 
a function of $|\alpha|$ for noise parameters 
$\eta=0.9,0.7,0.5,0.3,0.1$ for top to bottom curves.}
\end{figure}
%%%%%%%%%%%%%%%%%%%%%%%%%%%%%%%%%%%%%%%%%
For comparison, we consider the biphoton Bell state:
\begin{equation}
|\Psi_{2(p)}\rangle_{AB} ={1\over\sqrt2}
        ( |\updownarrow\rangle_A |\leftrightarrow\rangle_B
         -|\leftrightarrow\rangle_A |\updownarrow\rangle_B ),
\end{equation}
where $\{|\updownarrow\rangle,|\leftrightarrow\rangle\}$ denotes
single photon polarization directions.
After passing through the same lossy
channel, Alice and Bob share the state
\begin{equation}
\rho_{AB}^{\rm pol}=
\eta|\Psi_{2(p)}\rangle_{AB} \langle\Psi_{2(p)}|
+(1-\eta)\frac{1}{2}I_A\otimes|0\rangle_B\langle0|
\label{biphoton_entangle}
\end{equation}
with
$I_A$ the identity on mode $A$.
As a result, the fully entangled fraction is
\begin{equation}
f(\rho_A^{\rm pol})=\eta
\end{equation}
in that case, which is clearly less than for entangled coherent 
states with sufficiently small amplitudes (see Figure 1). 
Finally, we note that in \cite{enk} we give the analogous 
decoherence properties for a symmetric noise channel, 
describing the case where both Alice's and Bob's mode suffer losses.
\section{Concluding remarks}
In this paper, our effort was devoted to clarify
several properties of entangled nonorthogonal states.
We constructed 4 entangled states that generalize the standard Bell states.

Two out of these 4 ``quasi Bell states'' possess less than one unit of 
entanglement, the other two, however, possess exactly one unit. 
The latter two states were shown to be more
robust against decoherence due to photon absorption than are bi-photon Bell states.
The most important remaining problem is the physical realization of
such states, which is discussed in a forthcoming separate paper \cite{enk}.

\section*{Acknowledgment}
We are grateful to C.H.Bennett, S.M.Barnett, C.A.Fuchs,
A.S.Holevo, R.Jozsa,\\
M.Sasaki, and P.Shor
for helpful discussions.
The first idea of quasi Bell state was presented in QCM
$\&$ C-2000 held in Capri. \\
This work was supported by
the project in CREST Japan.

\end{document}